\documentclass[10pt,conference]{IEEEtran}
\IEEEoverridecommandlockouts

\usepackage{url}
\usepackage{cite}
\usepackage{amsmath,amssymb,amsfonts}
\usepackage{algorithmic}
\usepackage{graphicx}
\usepackage{textcomp}
\usepackage{xcolor}
\newcommand{\linebreakand}{%
  \end{@IEEEauthorhalign}
  \hfill\mbox{}\par
  \mbox{}\hfill\begin{@IEEEauthorhalign}
}
\newcommand{\rqn}[1]{RQ\textsubscript{#1}}
\newcommand{\rp}{\footnote{\url{https://doi.org/10.5281/zenodo.14209193}}}
\def\BibTeX{{\rm B\kern-.05em{\sc i\kern-.025em b}\kern-.08em
    T\kern-.1667em\lower.7ex\hbox{E}\kern-.125emX}}
\begin{document}

\title{
Automating Technical Debt Management: Insights from Practitioner Discussions in Stack Exchange

\thanks{This study was financed in part by the Brazilian Federal Agency for Support and Evaluation of Graduate Education (CAPES) - Finance Code 001. \\ 
This work was funded by FAPESP (2023/00488-5) and CNPq (313245/2021-5).}
}

\author{
    \IEEEauthorblockN{João Paulo Biazotto}
    \IEEEauthorblockA{
        University of Groningen \\ 
        Groningen, The Netherlands \\
        University of São Paulo\\
        S\~ao Carlos - SP - Brazil \\
        j.p.biazotto@rug.nl
    }
    \and
    \IEEEauthorblockN{Daniel Feitosa}
    \IEEEauthorblockA{
        University of Groningen \\
        Groningen, The Netherlands \\
        d.feitosa@rug.nl
    }
    \and
    \IEEEauthorblockN{Paris Avgeriou}
    \IEEEauthorblockA{
        University of Groningen \\
        Groningen, The Netherlands \\
        p.avgeriou@rug.nl
    }
    \and
    \IEEEauthorblockN{Elisa Yumi Nakagawa}
    \IEEEauthorblockA{
        University of São Paulo\\
        São Carlos-SP, Brazil \\
        elisa@icmc.usp.br
    }
}
\maketitle

\begin{abstract}
Managing technical debt (TD) is essential for maintaining long-term software projects. Nonetheless, the time and cost involved in technical debt management (TDM) are often high, which may lead practitioners to omit TDM tasks. The adoption of tools, and particularly the usage of automated solutions, can potentially reduce the time, cost, and effort involved. However, the adoption of tools remains low, indicating the need for further research on TDM automation. To address this problem, this study aims at understanding which TDM activities practitioners are discussing with respect to automation in TDM, what tools they report for automating TDM, and the challenges they face that require automation solutions. To this end, we conducted a mining software repositories (MSR) study on three websites of Stack Exchange (Stack Overflow, Project Management, and Software Engineering) and collected 216 discussions, which were analyzed using both thematic synthesis and descriptive statistics. We found that identification and measurement are the most cited activities. Furthermore, 51 tools were reported as potential alternatives for TDM automation. Finally, a set of nine main challenges were identified and clustered into two main categories: challenges driving TDM automation and challenges related to tool usage. These findings highlight that tools for automating TDM are being discussed and used; however, several significant barriers persist, such as tool errors and poor explainability, hindering the adoption of these tools. Moreover, further research is needed to investigate the automation of other TDM activities such as TD prioritization.
\end{abstract}


\section{Introduction}
\label{sec:intro}
Incurring technical debt (TD is not inherently harmful, as it can accelerate development or reduce costs in the short term; however, unmanaged TD can compromise a project's long-term maintainability and evolvability~\cite{Junior2022}. The most recent Stack Overflow Annual Developer Survey\footnote{\url{https://shorturl.at/Paq2K}} reveals that the accumulation of TD is a primary concern for developers, contributing significantly to frustration. Therefore, the quality and sustainability of software projects, as well as the motivation and productivity of developers, to some extent, depend on effective TD management (TDM)~\cite{Avgeriou2023}.

A key challenge related to TDM is the significant time required to carry out activities, such as TD identification and repayment. According to Besker et al.~\cite{Besker2019}, developers spend approximately 23\% of their time dealing with TD. This time is typically spent on additional tasks due to the existence of TD, such as additional testing and refactoring, ultimately increasing developers' workloads and reducing their productivity. This subsequently leads to higher software development costs~\cite{Besker2019}. TDM automation has then emerged as a solution to reduce the overall effort required for TDM, enabling continuous TDM throughout the development life cycle~\cite{Khomyakov2019}.

In principle, increasing the adoption of tools in TDM 
can lead to more automated and thus efficient processes to deal with TD; however, tool adoption proves rather challenging. Recent surveys with practitioners showed that they tend to use general-purpose tools (e.g., Jira)~\cite{Martini2018}. While such tools might benefit teams that do not perform TDM at all~\cite{Rios2020}, they do not provide TD-related features (e.g., measuring principal or interest of TD); consequently, they do not necessarily mature TDM practices in development teams~\cite{Besker2018}.

We argue that further research is necessary to understand the challenges of TDM automation. This can help to define better automated solutions and reduce the time and effort involved in managing TDM. In this study, we seek to understand which TDM activities practitioners are discussing w.r.t automation of TDM, what tools they report for automating TDM, and what challenges they face that require automation solutions. To this end, we conducted a mining software repositories (MSR) study focused on the Stack Exchange network\footnote{\url{https://stackexchange.com/}}.

Our study makes three key contributions to the field of TDM automation. First, we systematically analyzed discussions from the Stack Exchange network to identify the most commonly discussed TDM activities. We also highlight activities with limited evidence (e.g., prioritization and prevention). Second, we provide a comprehensive catalog of tools for TDM automation, serving as a practical guide for researchers and practitioners. By presenting the strengths and weaknesses of commonly discussed tools, the study helps practitioners make informed decisions when selecting TDM tools. We also identified tool usage trends, such as the most frequently mentioned tools, and linked these tools to specific TDM activities. This analysis  revealed under-supported activities where tool development is lacking. Finally, we identified and categorized the challenges practitioners face in TDM automation into two primary areas. The first category encompasses drivers for automation, including challenges like poor resource allocation and tight deadlines. The second category involves tool-related challenges, such as tool errors and lack of explainability, which hinder the effective adoption and use of automation solutions. 

The remainder of this paper is organized as follows: Section~\ref{sec:rw} discusses the related work and compares it to our results, highlighting our contributions.  Section~\ref{sec:sd} describes the methods used in this study, including data source selection, data collection pipeline, and data analysis approaches. The results are presented in Section~\ref{sec:results} and discussed in Section~\ref{sec:discussion}. Section~\ref{sec:tov} outlines the threats to the validity of this study and the actions taken to mitigate them. Finally, Section~\ref{sec:conclusion} presents the conclusions of this study and discusses potential future research directions.

\section{Background and Related Work}
\label{sec:rw}

Various activities have been proposed to aid practitioners in effectively managing TD~\cite{Alves2016, Guo2011, McGregor2012, Santos2013}. Li et al.~\cite{Li2015} summarized nine critical activities described in the literature. During \textbf{identification}, TD items are detected using techniques such as manual inspection or static code analysis. These items can be documented during the \textbf{representation and documentation} activities, and stakeholders are informed about TD items during the \textbf{communication} activity. TD items are monitored in the \textbf{monitoring} activity, ensuring that unresolved TD items are under control. The \textbf{measurement} activity quantifies the amount of TD in a system, enabling the \textbf{prioritization} activity, where TD items are ranked for resolution. These items can then be addressed during the \textbf{repayment} activity, which also deals with the issues caused by accumulated TD. Additionally, the \textbf{prevention} activity seeks to avoid unwanted TD from arising.

Previous work has investigated such activities in different contexts, including how they could be automated to make TDM more efficient and how these activities and TD in general are discussed on the Stack Exchange network~\cite{Alfayez2023,Kozanidis2022, Santos2022, Edbert2023, Peruma2021, Gama2020}. Such studies are closely related to ours and are also discussed in this section. First, in an observational study on the usage of the term ``technical debt" on the Stack Exchange network, Alfayez et al.~\cite{Alfayez2023} found 578 TD-related questions, which were categorized into 14 different categories, such as TD tools, TD repayment, TD representation, and TD definitions. The study identified 636 unique tags for TD-related questions, with ``SonarQube'' being the most used tag, followed by ``technical debt'', ``java'', ``agile'', and ``scrum'' among the top-10 most used tags. Furthermore, the findings showed that the most challenging questions for users. Furthermore, the findings showed that the most challenging questions for users were those classified under three categories---TD consequences, incurring TD, and TD tools---and that most questions were related to what to do with TD tools..

Kozanidis et al.~\cite{Kozanidis2022} conducted a study to understand how users request support with respect to TD. After reviewing and analyzing 415 questions from the Stack Exchange websites, the authors found that TD in architecture, code, and design are the most referenced TD types on Stack Overflow. Moreover, predictive models can accurately detect and classify TD questions and their urgency (urgent or not urgent), but fail to identify TD types mentioned in the question. Finally, they found that most questions present some degree of urgency, while TD repayment and TD management are the most recurrent themes.

Other research studies focused on more specific contexts (e.g., agile or security) to analyze  questions about TD. One such study by Santos et al.~\cite{Santos2022} compiled and analyzed 79 TD discussions on agile software development (ASD) from Stack Exchange Project Management. They pointed out eight types of TD in the context of ASD and identified 51 indicators of ASD-related TD, for instance, poorly written code, design problems, and bug occurrence. They found that the most commonly discussed TD types are process and people debt and that product owner and development team are the most important roles concerning ASD-related TD. Another study, which researched the scope of TD in security questions found on Stack Overflow, identified that 38\% of the analyzed 117.233 questions were security-related TD questions~\cite{Edbert2023}.

An empirical study by Peruma et al.~\cite{Peruma2021} found that Code Optimization, Architecture and Design Patterns, Unit Testing, Tools and IDEs, and Database are the top-five topics most associated with discussions about refactoring, a common practice for TD repayment.

A study by Gama et al.~\cite{Gama2020} investigated how developers commonly identify TD items in their projects. They found that Stack Overflow users commonly discuss the identification of TD, revealing 29 different low-level indicators to recognize TD items in code, infrastructure, architecture, and tests. They grouped low-level indicators based on their themes, producing a set of 13 high-level indicators, such as the presence of bad coding and the lack of good design practices. In addition, they classified all low- and high-level indicators into three different categories (Development Issues, Infrastructure, and Methodology) according to the type of debt that each of them was intended to identify.

The aforementioned studies show that practitioners often use Stack Exchange forums to discuss TD and TDM. However, these studies, at best, focus on listing the tools mentioned in these discussions without exploring them in depth.
Our study investigates the challenges of automating TDM in practice through existing tools or desired TDM features. Our findings contribute to TDM research and practice by exploring limitations in current tools and offering insights for researchers and tool developers to improve or create more effective solutions. Together, these contributions can lead to more efficient TDM processes and reduce the time required to manage TD.

\section{Study Design}
\label{sec:sd}
This section presents the objective and research questions (RQs) of this study, the data collection process, and how we analyzed the data to answer the RQs.

\subsection{Objective and Research Questions}
\label{sec:objective}
Considering the problems and gaps we presented in the two previous sections, the objective of this study, structured according to the Goal-Question-Metric template~\cite{vanSolingen2002}, is to %
\textit{``analyze \textbf{discussions} for the purpose of \textbf{identifying TDM activities, tools, and challenges} with respect to \textbf{TDM automation} from the point of view of \textbf{practitioners} in the context of \textbf{the Stack Exchange network}.''} To achieve this objective, we defined the following RQs:

\begin{itemize}

\item \textbf{\rqn{1} - What TDM activities do practitioners discuss in relation to TDM automation?} \\
While previous literature presents various tools for supporting TDM activities~\cite{Silva2022, Junior2022, Biazotto2023} (such as identification, measurement, and prioritization of TD items - see Section~\ref{sec:rw} for a full list), it remains unclear which specific TDM tasks practitioners focus on automating. This RQ addresses this gap by identifying the TDM tasks mentioned in discussions about TDM automation. The results of this RQ have the potential to impact TDM automation, helping researchers and tool vendors direct their efforts toward TDM activities that are most relevant to practitioners.

\item \textbf{\rqn{2} - What tools are reported as options for TDM automation?} \\
Existing literature on TDM tools~\cite{Silva2022, Junior2022, Biazotto2023} reports and evaluate tools that support one or more tasks of TDM. To complement such literature, an analysis of the usage of such tools in practice, specifically highlighting which tools are preferred by practitioners for TDM automation, is still necessary. This RQ investigates which tools are reported in the discussions about TDM automation. The results of this RQ can help to shape and improve tool adoption, leading to more efficient TDM.

\item \textbf{\rqn{3} - What challenges do practitioners report in discussions about TDM automation?} \\
This RQ analyzes the challenges reported by practitioners from two perspectives. First, we examine the challenges that drive practitioners to automate TDM activities (e.g., performing TDM activities more efficiently). Second, we identify the challenges practitioners encounter while automating TDM activities (e.g., the need for integration with other tools). The first perspective can support researchers in proposing tooling approaches that address these challenges, leading to more efficient TDM. The second perspective highlights problems that could inspire tool improvements, potentially increasing the adoption of such tools.

\end{itemize}

\subsection{Data Collection and Analysis}
\label{sec:data-collection}

\begin{figure*}
    \centering
    \includegraphics[width=\textwidth]{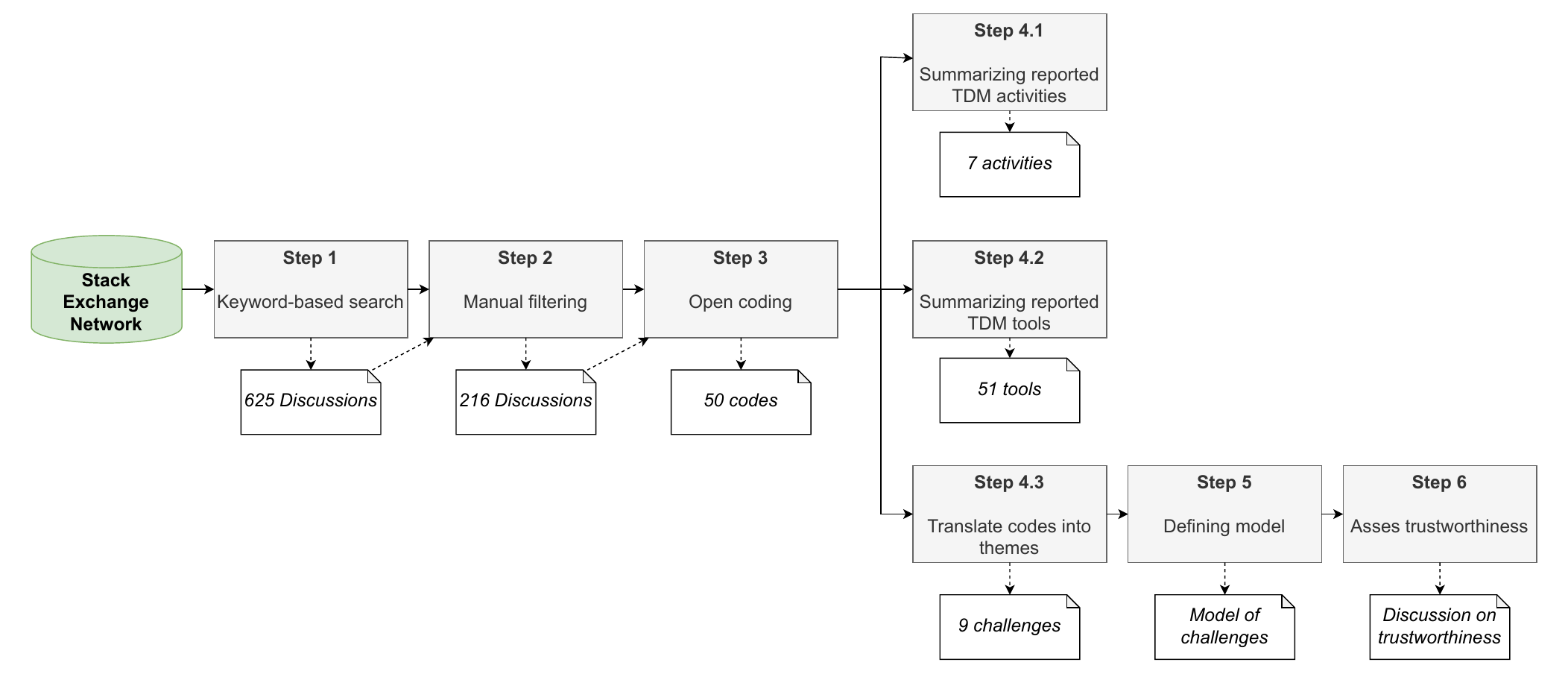}
    \caption{Data collection and analysis process}
    \label{fig:research-method}
\end{figure*}

Figure~\ref{fig:research-method} depicts the steps we followed to collect and analyze data in our study. The discussions' dataset was obtained from three Stack Exchange Q\&A websites: Project Management\footnote{\url{https://pm.stackexchange.com/}} (PM), Software Engineering\footnote{\url{https://softwareengineering.stackexchange.com/}} (SE), and Stack Overflow\footnote{\url{https://stackoverflow.com/}} (SO). These websites were chosen because they contain questions on specific programming problems, software tools commonly used by programmers, and software development methods and practices. Furthermore, two previous studies on TD discussions in the Stack Exchange (by Alfayez et al.~\cite{Alfayez2023} and Kozanidis et al.~\cite{Kozanidis2022}) successfully collected empirical evidence on the TD phenomenon utilizing these three Q\&A websites; this suggests that they are suitable data sources for our study.

\textbf{Project Management (PM)} is a Q\&A website for project managers and academics. It was established in 2011 and covers several topics, including project management frameworks, how to use tools to address problems linked to project management, and the profession or practice of project management ~\cite{Alfayez2023}.

\textbf{Software Engineering (SE)} is a Q\&A website for practitioners, academics, and students working within the software development life cycle. It was established in 2010 and focuses on topics relating to software development methods and practices, requirements, architecture, design, quality assurance, testing, configuration management, building, release, and deployment. 

\textbf{Stack Overflow (SO)} is a Q\&A website for programmers, academics, and students. It was established in 2008 and addresses questions that cover specific programming problems, software algorithms, and software tools commonly used by programmers. It is the biggest, well-known, and established Stack Exchange website~\cite{Alfayez2023}.

For data collection, 7z files were downloaded from the Stack Exchange Data Dump\footnote{\url{https://archive.org/details/stackexchange}}. The dump contained data up until 2024-04-02. An existing set of Python scripts provided by Feitosa et al.~\cite{Feitosa2024} was used for the automated data extraction (\textbf{Step 1} in Figure~\ref{fig:research-method}). The original scripts can search for a given keyword in the tags of each Stack Overflow question, and tie each question to its corresponding set of answers, comments, and post history. We extended the scripts to search for a given keyword also in the title and body of each question. We used the keywords ``technical debt'' and ``tech debt'' to search the discussions. Our decision to use these two keywords is based on evidence presented in previous studies~\cite{Gama2020,Kozanidis2022,Alfayez2023}, which show that these keywords are popular when people are discussing the TD phenomenon on Stack Exchange websites. The initial keyword-based search resulted in a set of 625 discussions.

The TD-related questions related to each discussion were manually examined to ensure the relevance of each discussion, i.e., whether a discussion was related to TD and discussed tooling or automation of TDM (\textbf{Step 2} in Figure~\ref{fig:research-method}). To guide this filtering, we defined inclusion and exclusion criteria, as follows:

\begin{itemize}
    \item \textbf{\textit{IC1}} - The discussion is related to tooling or automation in TDM.
    \item \textbf{\textit{IC2}} - The discussion implicitly suggests or presents a TDM practice that could be supported by tools or automated.
    \item \textbf{\textit{EC1}} - The discussion is not centered on TD.
\end{itemize}

A discussion was accepted if it matched at least one of the IC and did not match the EC. For instance, consider the following question: ``\textit{In real world, business initiatives always take higher priority as there are associated ROIs and deliver something tangible to the users. But there are technical initiatives and projects that need to be done to keep up with the different versions of software, upgrading to a newer platforms, architecture re-factoring etc. How can we plan, prioritize and manage such competing initiatives? Is there a model to quantify technical debt and its impact to the business?}'' In this case, the user is concerned about how to quantify TD (IC2) and its impact on the business, so this questions was considered relevant for our study.
In contrast, consider this question: ``\textit{I hear a lot of terms which aren't well known amongst programmers (or perhaps the ones I work with at work aren't very good apart from a few), such as ``technical debt" (which I studied and even see first hand at work). What other obscure/not-well-known terms are there? This is especially useful to know as interviewers sometimes mention complex terms and if I don't know what they mean, it can screw up the interview as it is in progress.}'' The main goal of this question was to discover ``obscure terms'' used by the software engineering community, and TD was just mentioned as an example (EC1). After  filtering, we remained with 216 discussions, which were further analyzed to answer the RQs.

We relied on thematic analysis to answer our RQs. Thematic analysis involves five main steps~\cite{Cruzes2011}: extract data, code data, translate codes into themes, create a model of higher-order themes, and assess the trustworthiness of the synthesis. We noted that each RQ had a different scope; for \rqn{1} and \rqn{2}, we considered only the two first steps of thematic analysis (i.e., extract data and code data), while for \rqn{3} we applied all the five steps. This is because \rqn{1} and \rqn{2} focus on listing activities and tools mentioned in discussions, so the codes we used are name of tools and activities. In contrast, \rqn{3} focuses on identifying challenges (themes) in TDM automation, and how practitioners deal with them. 

For extracting data, we collected 625 discussions from Stack Exchange (\textbf{Step 1} in Figure~\ref{fig:research-method}). After applying the inclusion/exclusion criteria  (\textbf{Step 2} in Figure~\ref{fig:research-method}),  216 discussions were selected to be further analyzed.

For coding the data, we adopted an approach based on grounded theory. This means that our coding approach includes open, axial, and selective coding, which were applied in all the 216 selected discussions. Open coding was applied to break down the discussions into \textit{indicators}, a piece of evidence present in the data that was analyzed to check any pattern in the dataset (\textbf{Step 3} in Figure~\ref{fig:research-method}). 

To answer \rqn{1}, we combined the list of TDM activities summarized by Li et al.~\cite{Li2015} as codes with the codes generated from the data. We then summarized the activities and used descriptive statistics to analyze the prevalence of each one (\textbf{Step 4.1} in Figure~\ref{fig:research-method}). We followed a similar approach to answer \rqn{2}, but we used post-formed codes only (i.e., the name of each tool we identified); descriptive statistics was used to present the most cited tools (\textbf{Step 4.2} in Figure~\ref{fig:research-method}). 
To answer \rqn{3}, we applied the remaining three steps of thematic analysis, as follows:

\begin{itemize}
    \item \textbf{Translate codes into themes (Step 4.3 in Figure~\ref{fig:research-method}):} The theme emerged through the process of coding, categorization, and reflective analysis. They represented abstract constructs that imbued recurring experiences with meaning and coherence. This step involved both axial and selective coding and emphasized constant comparison, which aided in mitigating potential biases and enhancing data exploration;
       
    \item \textbf{Create a model of higher-order themes (Step 5 in Figure~\ref{fig:research-method}):} The themes that emerged in the previous step were further explored and interpreted to create a model consisting of higher-order themes and relationships between them. To answer \rqn{3}, we proposed a model that captures the main challenges that practitioners have in TDM, and how tool adoption is used to deal with such challenges; and
        
    \item \textbf{Assess the trustworthiness of the synthesis (Step 6 in Figure~\ref{fig:research-method}):} Assessing the reliability of interpretations derived from thematic synthesis involved constructing arguments supporting the most plausible interpretations. To evaluate the reliability of the results, we examined the concepts of credibility, confirmability, dependability, and transferability as elaborated in Section~\ref{sec:tov}.
\end{itemize}

\section{Results}
\label{sec:results}

\subsection{Demographics}

\begin{figure} [!ht]
    \centering
    \includegraphics[width=0.4\textwidth]{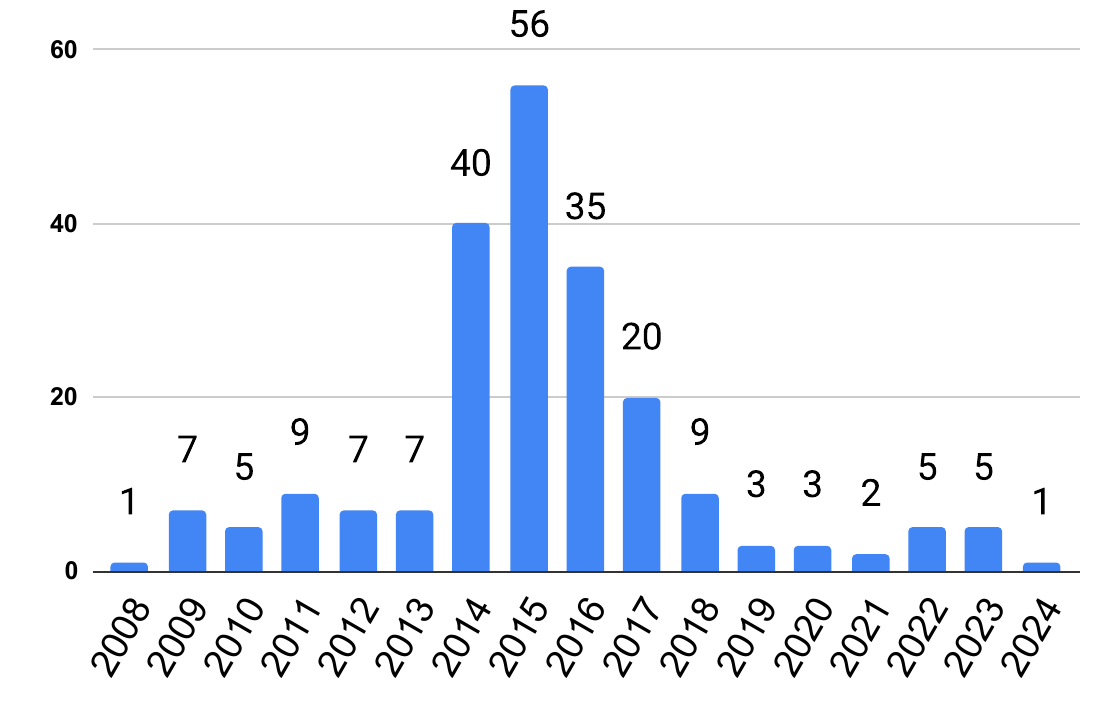}
    \caption{Number of discussions per year}
    \label{fig:discussions-per-year}
\end{figure}

We found 216 discussions related to tooling for TDM, with at least one post per year from 2008 to 2024, as shown in Figure~\ref{fig:discussions-per-year}. The yearly distribution of discussions shows a marked increase in interest between 2014 and 2017, during which approximately 70\% of the discussions were initiated. Notably, since 2019, the number of discussions has decreased significantly. This decline might be due to a number of reasons, including improvements in tools and recurrent problems in tools that are already discussed in previous questions. Besides, the growing adoption of large-language-model-based systems, such as ChatGPT that has become widely used as a Q\&A tool among software engineers after 2022, has likely also contributed to the decrease of questions about tooling for TDM in Stack Overflow~\cite{Kabir2024}.

\begin{figure}
    \centering
    \includegraphics[width=0.3\textwidth]{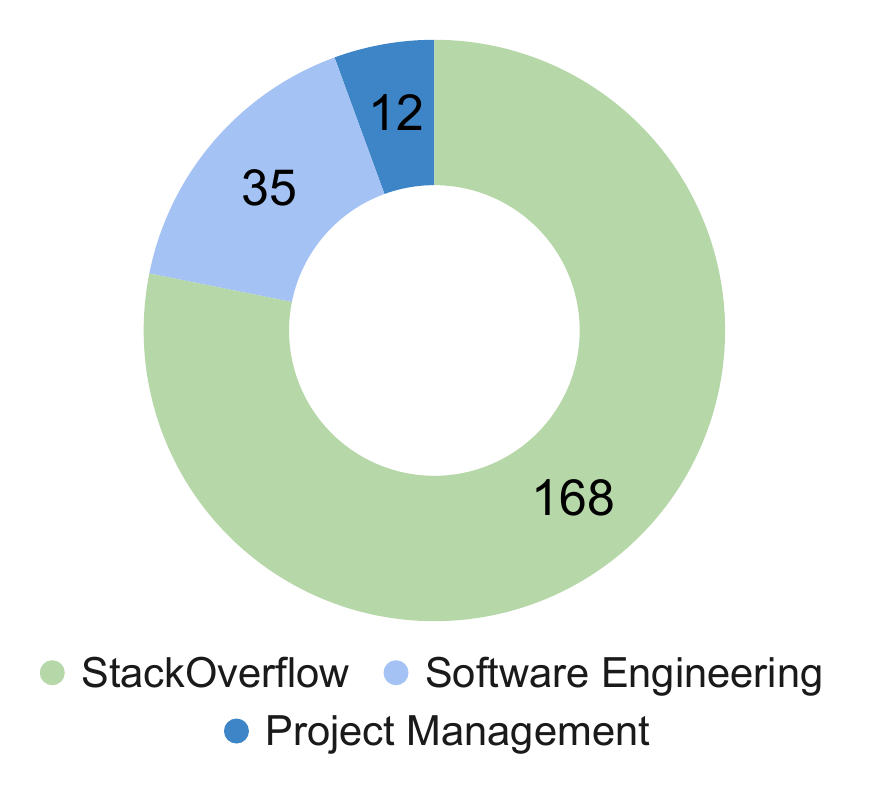}
    \caption{Number of discussions per website}
    \label{fig:discussions-per-source}
\end{figure}

Regarding the number of discussions retrieved from each website (Figure~\ref{fig:discussions-per-source}), Stack Overflow contributed the most, followed by the Software Engineering and Project Management Stack Exchanges. This result was expected, given that Stack Overflow is the most widely used and well-known website on Stack Exchange. Nonetheless, we observed that discussions from Software Engineering and Project Management bring a unique perspective on TDM tooling---particularly from a management
viewpoint---which enriched the results and discussion in this study.

\subsection{\rqn{1} - TDM Tasks}
\label{results-rq1}

Table~\ref {tab:tdm-activities} presents the list of identified activities (see Section~\ref{sec:rw} for an explanation of each activity) and the number of discussions highlighting each activity from the total of 216 discussions. It is worth noting that a small number of discussions address more than one task, so the sum of discussions in the table is higher than the number of selected discussions. Further in this section, we also examine the aspects of TDM automation that practitioners are most concerned with, considering each activity. 

\begin{table}[!ht]
\centering
\caption{Number of discussions related to each TDM activity}
\label{tab:tdm-activities}
\begin{tabular}{lr}
\hline
\textbf{Activity}  & \textbf{\# discussions} \\ \hline
Identification & 95                           \\
Measurement    & 43                           \\
Payment        & 39                           \\
Documentation  & 23                           \\
Monitoring     & 12                           \\
Prioritization & 10                            \\
Prevention     & 6                            \\ \hline
\end{tabular}
\end{table}

Regarding \textbf{identification}, users primarily sought tools to analyze source code. For example, one user stated that \textit{``I want to determine the readability of the code written by an author. Thus, I am looking for a tool compatible with Python that will provide me with such functionality.''} From a more technical perspective, practitioners were also interested in integrating tools for identifying TD within existing development tools, such as continuous integration (CI) managers like Jenkins and Travis CI: \textit{``We use Continuous Integration and Continuous Delivery, so we run SonarQube analysis for every CI build to provide immediate feedback to developers. This ensures that issues are identified before the end of the sprint and prevents the accumulation of new technical debt.''}

Discussions related to TD \textbf{measurement} primarily focused on (a) finding automated solutions to calculate the accumulation of TD or (b) understanding how existing tools calculate the amount of TD. For (a), users often sought ways to obtain an overview of project quality, as one user asks: \textit{``Is there a model to quantify technical debt and its impact on the business?''} Regarding (b), users who already employ tools like SonarQube and Checkstyle sought clarification on how these tools measure TD to make more informed decisions, as illustrated by another user: \textit{``How does SonarQube measure maintainability issues?''}

In discussions related to \textbf{payment} of TD, users primarily sought alternatives to assist with three tasks: (i) automated test generation (to deal with the lack of tests), (ii) rewriting source code, and (iii) refactoring source code. For automated test generation, users looked for solutions to address gaps in testing: \textit{``In other words, I have high-level tests, but I would like to automatically create low-level unit and integration tests based on the execution of these high-level tests.''} When rewriting code, users expressed concern about the effort required and sought tools to make the process more efficient: \textit{``How can I efficiently approach this rewrite?''} Finally, a significant concern that users reported about refactoring was the challenge of allocating time for it while simultaneously developing new features: \textit{``When to refactor and when to extend while accruing technical debt?''}

When it comes to automating the \textbf{documentation} of TD, the primary concern was how to use tools like issue tracking and backlogs to maintain a registry of TD items, as one user explained: \textit{``I need to record each item outside of a specific iteration to ensure that it is visible and easily-reported all the time.''} Additionally, users sought tools to help visualize and present TD information effectively, as shown by another user's question: \textit{``How do you visually display your results to stakeholders and management?''}

The users were also seeking automated solutions to help the \textbf{monitoring} and tracking of TD items. Specifically, they wanted tools that keep them aware of outstanding TD items, as one user notes: \textit{``I've found that making time to go back and fix them, let alone remembering the full list of to-do items, can be challenging at best. Can you recommend tools, resources, or tricks that help you effectively manage [monitoring] your technical debt?'' }Additionally, users were looking for advice on better using specific tools for tracking TD, such as \textit{``We are considering using NDepend to start tracking some of our technical debt, particularly around hard-to-maintain methods and cyclomatic complexity.''}

\textbf{Prioritization} of TD items to be paid and how to use tools to support such decisions was also a concern: ``\textit{I'm looking for a way to quantify where my team should spend its time addressing technical debt in our codebase. One idea for this is to measure file churn (edits over time). I got the idea from this video where Michael Feathers talks about escaping the technical debt cycle}.''

Finally, the \textbf{prevention} of new TD items from being introduced into a project is another concern that leads users to seek automated solutions. As one user explains, \textit{``At the moment, I want to ensure that any new code checked in is evaluated and issues flagged [as TD].''} Additionally, users are looking for tool support for code review, so they could save time in this process and prevent problematic code from entering the codebase. One user notes, \textit{``Nowadays, I try to review my teammates' check-ins when time allows (though it is rare), but there is no automated process to prevent unwanted changes.''}

\subsection{\rqn{2} - TDM Tools}

A total of 51 tools were reported in the selected discussions.
Due to space limitations, Table~\ref{tab:tdm-tools} lists the 10 most frequently mentioned tools; the full list of all 51 tools can be found in our replication package\rp.

\begin{table}[!ht]
\centering
\caption{Top-10 most cited tools and the number of discussions mentioning each tool}
\label{tab:tdm-tools}
\begin{tabular}{lr}
\hline
\textbf{tool} & \multicolumn{1}{l}{\textbf{\# discussions}} \\ \hline
SonarQube     & 138                                         \\
SQUALE         & 10                                          \\
Checkstyle    & 5                                           \\
JaCoCo        & 5                                           \\
FxCop         & 4                                           \\
FindBugs      & 3                                           \\
Cppcheck      & 3                                           \\
NDepend       & 2                                           \\
ESLint        & 2                                           \\
Xdebug        & 2                                           \\
CodeMaid     & 2                                           \\
SonarSource   & 2                                           \\ \hline
\end{tabular}
\end{table}

Tools were explicitly mentioned in 202 out of 216 discussions. In these discussions, users referred to tools in two main scenarios: (a) seeking help to resolve an error in a tool already in use; or (b) suggesting a tool as a potential solution for a reported challenge in TDM. Most mentions appeared in discussions regarding tool errors (139 out of 202), while the remaining mentions were suggestions of a tool as a solution to a management challenge.

Overall, \textit{SonarQube} was cited in 138 discussions, representing over 80\% of the dataset. This high number of mentions confirms the trend of using \textit{SonarQube} for TDM automation. Other tools, such as \textit{SQALE}, \textit{CheckStyle}, and \textit{Jacoco}, were also mentioned by users, albeit much less frequently. Furthermore, 31 tools, including \textit{JDepend}, \textit{Cast}, and \textit{PMD}, were referenced only once. This outcome is somewhat surprising, given that these are well-known tools for TDM automation.
We note that while \textit{SonarQube} is heavily mentioned, most discussions (92\%) regarded problems with it. Meanwhile, other tools (e.g., \textit{PMD}) are reported only as solutions to challenges.

Table~\ref{tab:tdm-tools-tasks} reports the number of tools associated with discussions about each TDM activity identified in \rqn{1}. Most tools (29) were mentioned in the context of identification, followed by payment and measurement (10 each). Notably, no tools were reported in discussions on prioritization, indicating a potential lack of tool support or knowledge of existing tools for this activity. Furthermore, tools can be used in multiple activities; for instance, \textit{SonarQube} was reported to identify and measure TDM.

\begin{table}[!ht]
\centering
\caption{Number of tools related to each TDM activity}
\label{tab:tdm-tools-tasks}
\begin{tabular}{lr}
\hline
\textbf{Activity}  & \multicolumn{1}{l}{\textbf{\# of related tools}} \\ \hline
Identification & 29                                            \\
Payment  & 10                                            \\
Measurement   & 10                                            \\
Monitoring    & 7                                            \\
Documentation & 5                                             \\
Prevention  & 3 \\ \hline
\end{tabular}
\end{table}

\subsection{\rqn{3} - Challenges in automating TDM}
\label{sec:rq3}
\begin{figure*}[ht]
    \centering
    \includegraphics[width=\textwidth]{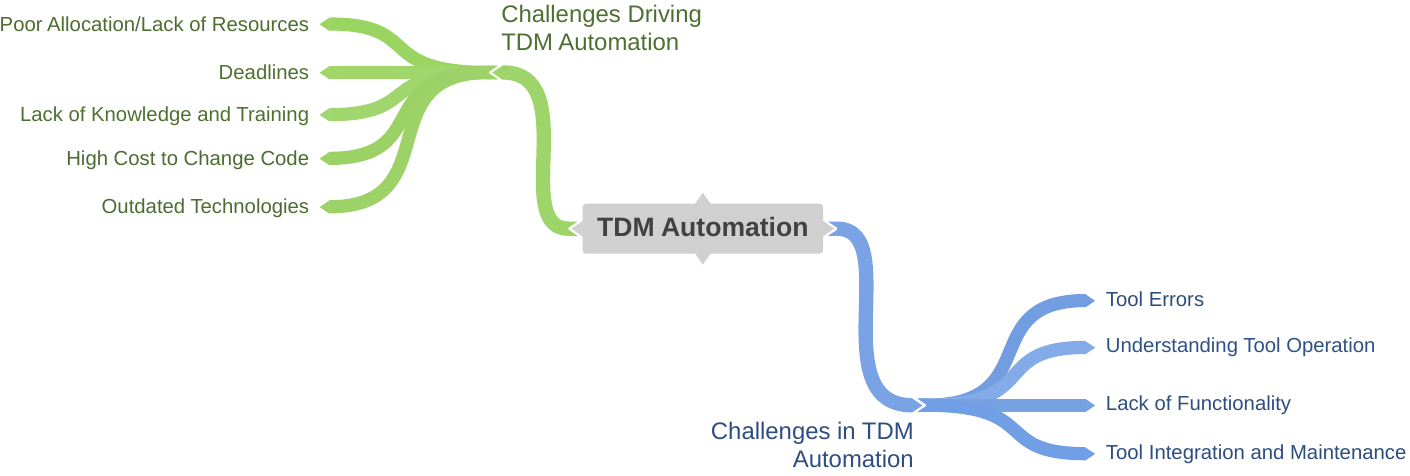}
    \caption{Challenges related to TDM automation}
    \label{fig:challenges}
\end{figure*}

To address \rqn{3}, we outlined the challenges that practitioners reported regarding TDM automation. These challenges can be divided into two main categories: (i) \textit{challenges that drive practitioners to automate TDM} refer to problems that compel practitioners to look for automated solutions for TDM (e.g., cost); and (ii) \textit{challenges that practitioners face while automating TDM} refer to problems in automated solutions already in place (e.g., tool errors). Figure~\ref{fig:challenges} summarizes the identified challenges in each category.

Regarding \textbf{challenges driving automation}, we observe that certain causes of TD also acted as key drivers for automating TDM. Practitioners sought automated solutions when these causes significantly impacted the project (e.g., a short deadline to be met, while TD must also be addressed). The challenges in this category are elaborated in the following paragraphs.

\textbf{Poor Allocation/Lack of Resources:} A main cause of TD was also one of the main reasons why practitioners looked for TDM automation through tools: ``\textit{There is a project with huge technical debt which is currently in the maintenance phase. However, because of the technical debt a lot of effort is needed to keep the thing running --- 3 people should constantly work on fixing bugs to deal with the workload}.'' In this case, the user was facing a lack of resources to manage TD and was looking for automated solutions that could help them deal with the workload of the project's maintenance. On a second example, practitioners looked for tool support when they needed to balance the development of new features and the payment of TD: \textit{``How should we balance these two things so that the developers still feel they can make progress on the engineering initiatives while remaining focused on finishing the feature set?''}. In such cases, they looked for tools that could help them find a balance between developing features and managing TD within their resource constraints. Finally, they also looked for tools that can support their planning and allocation of time for code quality management or refactoring: ``\textit{Where is the line of caring about code quality enough (not too less and not too much)?}''

\textbf{Deadlines:} Automation was also intended when the time to deal with TD was short, and deadlines must be met: ``\textit{I'm on a tight deadline and finding it harder and harder to write good, DRY, SOLID, code. It's becoming more enticing to copy/paste chunks of code to make slight variations to its behavior as design time goes up. It's also taking a long time to get back into the code base whenever I have to do a context switch (From one project then back to this one), I have a feeling of dread whenever I go back to work on this project}.'' This urgency leads practitioners to seek automated solutions that can help  rewrite code and meet the deadlines while still ensuring code quality.

\textbf{Lack of Knowledge and Training:} Lack of knowledge on the team was also a cause of TD that led to the need for automation: ``\textit{His programming background is known for his low quality. On his first experiences on the team, he demonstrated a lack of a clarity on what he was doing, and if he keeps work like this he'll increase the project's TD}.'' Practitioners looked for tools that can overcome and prevent errors from less experienced developers, so that they can ensure higher quality on the project. In the same direction, they also looked for tools to overcome lack of training: ``\textit{In an ideal world, each new joiner is expected to be trained with the practices and standards applied on the project. Happens that the ideal world is far from reality. When project starts getting off track, is not rare to have more manpower added on it, which is doomed to failure}.'' In such cases, tools can be used to support newcomers during development and prevent them from incurring TD.

\textbf{High Cost to Change Code:} The high cost associated with modifying or extending legacy code was also reported as a driver to TD automation: \textit{``Our team has inherited a very large and brittle python 2 (and C,C++, few others) codebase that is very difficult and costly to update. Tons of dependencies. Very few tests. Adding behavior improvement and converting to python 3 both have appeared to be monumental tasks. Even making small changes for a new release we've had to revert many times as it's broken something.''}. Subsequently, the team asked for automated dependency analysis and code visualization tools to help reduce the cost of code change.

\textbf{Outdated Technologies:} Another driver of automation was how to manage outdated technologies on projects: ``\textit{Unfortunately, as we were making progress with deploying said application, the DevOps team Management Stopped us because they said that .NET 3.1 is No longer supported by Microsoft. Therefore, they said that inspectors/auditors would Raise serious concerns about having Unsupported technologies deployed to production environment.}'' In this case, practitioners were looking for automated practices that can help them avoid these problems in the future: ``\textit{What kind of  management techniques/processes/procedure/tools could have been put in place in order to prevent said setback incident?}''

In the second category (i.e., \textbf{challenges in TDM automation}), practitioners reported issues with tools already in use, seeking support to resolve problems related to tool functionality (e.g., tool errors). The challenges in this category are elaborated in the following paragraphs.

\textbf{Tool Errors:} When tools were already in place, errors in such tools were the most reported challenges. In fact, 106/216 discussions showed problems related to specific tools: ``\textit{We have also been using SonarQube for analysis but have not had much luck with the preview or incremental modes and GitHub integration.}'' or \textit{``In analysis, I see that FXcop rule violations are captured. But when I browse to SonarDash board, It shows technical debt as 0 and issues as 0.''}. When practitioners face such errors, they look for help online. This category also includes problems practitioners report about tool configuration and usage: ``\textit{It looks SonarQube's ``History'' diagram is always configured for about a two-month period. I'd like to produce a diagram that shows decreasing technical debt for a specific time period that correlates with a certain clean-up initiative? How can I configure the diagram to show a specific time period?}'' 

\textbf{Understanding Tool Operation:} Another problem reported by practitioners was the lack of explainability of tools. Users in Stack Exchange frequently reported that they did not understand how a certain tool calculates or identifies TD: \textit{``How does SonarQube calculate its Maintainability Rating?''} or \textit{``My question is what is the logic behind SQALE to Technical Debt ratio mapping? Why SQALE A rating is Tech Debt in range from 0\% to 5\%. But not 0\% to 3\% for instance? How should I define a SQALE rating limits?''}

\textbf{Lack of Functionality:} When tools were already in place to support TDM, practitioners can also face a lack of features to perform certain tasks: \textit{``What we are missing is the same exact feature for COVERAGE KPIs, currently we are not given the option to click on a delta in the coverage tab and see files and associated changes, is it something available but we are not seeing ? is this sth coming soon? We also have version 5.6 and I did not find the option in there as well.''} or \textit{``I believe at present there are no such facility in sonar  , do we have any such plugin or other way to achieve this.''}

\textbf{Tool Integration and Maintenance:} Keeping tools within the development workflow was also challenging for practitioners: \textit{``I'm certain that I could get most of this using a hand-rolled linter running under Jenkins but I'd rather not have to maintain this. It's a nice thing to have not really a core part of the project I want to spend too much time on. In other words I want a ready to roll solution.''} Besides, they often faced problems in integrating such tools with other development tools: \textit{``Is there any way to convert OpenCover reports into the format of one of the aforementioned tools? Or, is there some other way to squeeze data from OpenCover into NDepend?''}

\section{Discussion}
\label{sec:discussion}

In this section, we revisit each RQ to discuss our findings and elaborate on their implication for practitioners and researchers.

\subsection{Interpretation of the results of \rqn{1}}
\label{sec:intepret-rq1}

The findings of \rqn{1} show that \textbf{identification} was the most frequently broached activity, suggesting that practitioners still struggled with the initial recognition of issues within TDM processes. This focus aligns with previous literature, which also emphasizes the importance of identifying TD as a critical first step in its management. Another interesting result is that practitioners were looking for alternatives to identify TD by integrating tools in the development environment (e.g., CI Managers). This confirms previous evidence provided by Junior and Travassos~\cite{Junior2022} and Biazotto et al.~\cite{Biazotto2023} pointing out to a need for more integration of tools and for embodying TDM within the development workflow.

Although the \textbf{measurement} and \textbf{payment} activities also received attention, they were brought up less frequently. While one may perceive this finding as ``practitioners not regarding these activities to the same level of importance,'' a more likely explanation is that practitioners had not reached that level of automation yet; in other words, you cannot measure what you cannot identify, let alone fix it. The same reason probably holds for the even lower mention of later activities such as \textbf{prioritization}. The activities have dependencies on each other: prioritization also depends on effective measurement, so to automate the former, you first need automation of the latter.

\subsection{Interpretation of the results of \rqn{2}}
\label{sec:intepret-rq2}

In \rqn{2}, we found 51 tools that were mentioned in the discussions, with SonarQube being the most mentioned (in 138 discussions). This finding aligns with previous evidence~\cite{Avgeriou2023,Avgeriou2020} and the overall perception of this tool's popularity. However, it is also worth highlighting that around 92\% of the discussions are about problems in the execution or configuration of SonarQube and its plugins. This indicates that improving ease of use (e.g., through better documentation) in tools is essential to better support practitioners and increase TDM tools' adoption.

This suggests two key points: (i) many users may need to be more familiar with other available tools, indicating a need to raise practitioners' awareness of and experience with these alternatives; and (ii) the relatively low frequency of reported errors in other well-known tools (e.g., Checkstyle) might explain why users did not discuss them on Stack Exchange, and why they were under-represented in our dataset.

Regarding the activities, identification is understandably the most commonly addressed, since this is one of the core activities supported by SonarQube.

We also note that payment was well-discussed, indicating that practitioners were actively looking for tools to reduce the accumulation of TD. However, there might be a severe lack of suitable tools for this activity, as a recent study showed that only 5\% of TDM tools addressed any form of payment~\cite{Biazotto2023}. On a similar note, no tools covering TD prioritization were mentioned in discussions, although at least a few examples of such tools exist~\cite{Diaz-Pace2020-bk,Nikolaidis_2021}, even if they are not as mature. For instance, recent literature has explored the use of genetic algorithms for selecting refactoring candidates~\cite{Maikantis2020-hr}, which can be a fair starting point. 

Although multiple tools for TDM have been cataloged in the literature~\cite{Avgeriou2020,Silva2022,Biazotto2023}, less than a half of these tools were mentioned in the discussions. Considering the evidence we have, we cannot conclude that we did not find mentions of them because practitioners did not face problems with these tools or due to a lack of awareness about the existence of such alternative tools. Thus, we argue that further research could be carried out to better understand the selection and usage of tools for TDM automation.

\subsection{Interpretation of the results of \rqn{3}}
\label{sec:intepret-rq3}

In \rqn{3}, we identified five key drivers for the automation of TDM. Notably, three of them can be directly linked to the causes of TD (i.e., poor allocation of resources, deadlines, and lack of knowledge and training). These findings complement existing literature on TDM automation, which primarily attributes the high costs associated with TDM as a motivator for automation. 

While addressing the challenges driving practitioners to automate TDM has the potential to improve software quality (e.g., dealing with high cost for code change), dealing with challenges related to automated solutions already in place (e.g., understanding tool operation) is equally important. In \rqn{3}, we found that around 60\% of the selected discussions address tool-related issues (particularly SonarQube). These findings raise a significant concern: tools limitations may hinder the adoption of TDM automation, potentially increasing development costs. Although this hypothesis requires confirmation in future studies, improving the quality and precision of existing tools is essential.

Another recurring issue practitioners reported was the need for more explainability in tools. The adoption of TDM tools depends on practitioners trusting their accuracy~\cite{Tsoukalas2024}. To build this trust, vendors must provide clear explanations of how the analyses are conducted. Additionally, making tools easier to use (e.g., by improving documentation) is necessary to reduce the number of errors practitioners encounter when installing and using these tools. 

\subsection{Implications for practitioners}
\label{sec:impl-practicitioners}

The results of this study offer valuable insights for practitioners in key aspects of TDM management. First, the findings can help practitioners enhance decision-making about their processes. For example, since resource allocation is a significant challenge in TDM and leads to automation, practitioners should  adopt and integrate tools in their workflow earlier on to prevent suboptimal resource allocation in the future. Additionally, we provide a list of used tools and rationales in a replication package~\rp, which can assist practitioners in selecting the most suitable ones.

We also advise tool vendors to explore ways to improve the explainability of tools. In particular, vendors could focus on providing reports or documentation to help practitioners make sense of the analysis results. Enhancing tool functionality and interoperability is also crucial. We recurrently noticed a desire for tool integration into existing workflows. Thus, vendors could strive for easier integrations, which could drive higher adoption.

\subsection{Implications for researchers}
\label{sec:impl-researchers}
The results of our study open several promising research directions. First, we collected evidence that identifying TD is the most commonly automated activity. However, other activities are under-represented, and research for  enabling and expanding automation for other TDM activities could be of merit. For example, while the prioritization of TD was mentioned in discussions  a few times (see \rqn{1}), there are only few and rather immature tools supporting this activity. There is a clear need to explore and propose automated approaches for TD prioritization.

Regarding tool explainability, it would be valuable to investigate how large language models (LLMs) could enhance these aspects. For instance, training models to read and summarize SQALE reports could increase practitioners' ease of digesting the information and drive higher adoption. Furthermore, although recent literature highlights researchers' active role in developing tools for TDM~\cite{Biazotto2023}, increasing the transfer of such tools to the industry is critical.

Our findings on the challenges related to automation reveal that causes of TD can also drive automation of TDM. This suggests that the consequences of TD accumulation impact not only the quality of the project but also the development processes and workflow themselves. Therefore, it may be interesting to extend research on TDM beyond code quality and, particularly, toward integrating tools and practices in development workflows and toolchains. Such work could provide valuable insights for shaping more effective automated approaches to managing TD.

\section{Threats to Validity}
\label{sec:tov}

As in any empirical study, there are several threats to the validity of our study. We organize such threats considering the guidelines presented in \cite{Wohlin2012,Guba1981,Cruzes2011}.

\textbf{Construct validity} (\textit{credibility} in the context of thematic synthesis~\cite{Guba1981}) regards the connection between the RQs and the study objects. The main threat for construct validity is the discussions were selected. To mitigate this threat, we defined a set of inclusion and exclusion criteria, which were rigorously followed for the selection of discussions. The accuracy of identifying TDM activities poses a potential threat, as these activities are sometimes only implied in discussions. To address this, we grounded our interpretation on a well-established set of activities defined by Li et al.~\cite{Li2015}.

\textbf{External validity} (\textit{transferability} in the context of thematic synthesis~\cite{Guba1981}) is concerned with the degree to which the findings can be generalized from the sample to the population. In our case, all the discussions were collected from the Stack Exchange network. To mitigate this potential bias, we chose three of the most used Stack Exchange websites, which were proven to be useful for looking for discussions about TD~\cite{Alfayez2023}. Nonetheless, it is not possible to ensure a balance on practitioner backgrounds, and we acknowledge our findings are deeply bound to our sample (as any other MSR study), and the results might not represent the entire population (i.e., software practitioners).

\textbf{Reliability} (\textit{confirmability} and \textit{dependability} in the context of thematic synthesis~\cite{Guba1981}) considers the bias from the researchers in data collection or data analysis. To mitigate bias from the interpretation of open-ended responses (i.e., challenges), we applied thematic synthesis, which is a well-established method for qualitative data analysis. We also applied an integrated approach for open coding, encompassing three steps. Specifically, the first author and an external collaborator coded all the discussions and discussed potential disagreements with the other authors. In the conflicting cases, the first three authors and the external collaborator discussed the codes until a consensus is achieved.

Although we used a grounded-theory-based approach to code the answers and define the themes (as presented in Section~\ref{sec:data-collection}), we did not follow all the steps of grounded theory. Specifically, the two steps (data collection and data analysis) were not concurrent (i.e., we collected all the data before analyzing it), preventing us from claiming the theoretical saturation of our challenges. Thus, replications of our study can help capture potential challenges not identified in \rqn{3} and move the model toward theoretical saturation. To support such replications and reproductions, we created a replication package\rp with all the necessary data and scripts to run the analyses.

\section{Conclusion and Future Work}
\label{sec:conclusion}

In this study, we reviewed discussions about TDM automation from the Stack Exchange network. Specifically, we found that identification was the most cited TDM activity and could be a good starting point to increase the automation of TDM. We also found 51 tools mentioned by practitioners for dealing with TD. SonarQube was heavily mentioned, but most discussions (92\%) regarded problems with it, indicating a clear need to increase the quality of tools for TDM.

We also found that adopting tools earlier in the development process can mitigate resource allocation challenges and reduce the long-term costs of managing TD. Furthermore, improving tool reliability, usability, and explainability are key for TDM automation. Additionally, developing clear documentation and enabling seamless integration into existing workflows are critical to driving adoption. 

In terms of research directions, expanding automation beyond TD identification, such as TD prioritization and prevention, is a critical research avenue. Furthermore, integrating emerging technologies like LLMs could improve tool functionality and explainability, enabling more effective TDM processes.

As future work, we plan to explore how prioritization of TD items can be effectively automated. Machine learning models or rule-based systems might be useful for analyzing historical data and ranking items by business impact, for instance. Prevention mechanisms, such as real-time feedback during code development, could also help avoid introducing new TD in projects, and we plan to investigate approaches that could support this.
Finally, we also plan to investigate how pipelines and toolchains could be integrated into the development workflow and enable higher levels of automation and integration.


\bibliographystyle{IEEEtran}
\bibliography{references}

\end{document}